\documentclass[twocolumn,twoside,slac]{revtex4}
\usepackage{graphicx}
\usepackage{fancyhdr}
\begin{document}

\title{HEP Applications on the Grid Canada Testbed}

\author{R.J. Sobie}
\affiliation{Institute for Particle Physics of Canada and 
Department of Physics and Astronomy,
University of Victoria, Victoria, British Columbia, Canada}

\author{A. Agarwal, J. Allan, M. Benning, R.V. Kowalewski,
G. Smecher, D. Vanderster, I. Zwiers}
\affiliation{Department of Physics and Astronomy,
University of Victoria, Victoria, British Columbia, Canada}

\author{G. J. Hicks}
\affiliation{
Common IT Services, Solutions BC,
Victoria, British Columbia, Canada}

\author{R. Impey, G. Mateescu}
\affiliation{Institute for Information Technology,
National Research Council of Canada, Ottawa, Ontario, Canada}

\author{D. Quesnel}
\affiliation{CANARIE Inc., Ottawa, Ontario, Canada}

\begin{abstract}

A Grid testbed has been established using resources at 12 sites across Canada 
involving  researchers from particle physics as well as other 
fields of science.  
We describe our use of the testbed with the BaBar Monte Carlo production and 
the ATLAS data challenge software.  
In each case the remote sites have no application-specific software stored 
locally and instead access the software and data via AFS and/or
GridFTP from servers located in 
Victoria.  
In the case of BaBar, an Objectivity database server was used for 
data storage.
We present the results of a series of initial tests of the Grid testbed 
using both BaBar and ATLAS applications.   
The initial results demonstrate the feasibility of using generic Grid 
resources for HEP applications.

\end{abstract}

\maketitle

\thispagestyle{fancy}


\section{Introduction}

One of the motivations for establishing a computational Grid is the
ability to exploit unused computing cycles at remote facilities.
In Canada, there has been a significant injection of new funding
for mid-size capability and capacity computational facilities. 
One of the conditions of the funding is that these facilities must
make 20\% of the resources available to outside users.
A number of these facilities are associated with high energy physics (HEP)
centers or groups.
However, the majority are facilities that are 
shared between many research fields.
We would like to exploit these unused cycles for HEP applications 
such as Monte Carlo simulations.
This paper describes the Grid Canada Testbed which consists of 
resources at both HEP and non-HEP sites.  
We demonstrate that HEP simulations can be run
without installing application-specific software at the remote site.
We discuss the results of a series of tests we have run over the Testbed
and conclude with a discussion of our plans for the future.

\begin{figure*}
\begin{center}
\includegraphics[width=100mm]{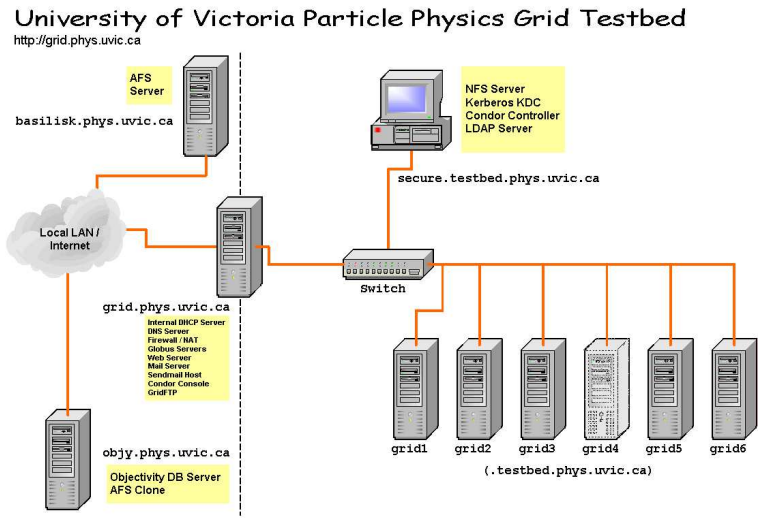}
\end{center}
\caption{
The University of Victoria Grid Testbed showing the AFS and 
Objectivity Servers as well as the head node (grid.phys.uvic.ca) 
and processing nodes (grid1, ..., grid6). 
}
\label{uvictb}
\end{figure*}

\section{Grid Canada Testbed}

Grid Canada (www.gridcanada.ca)
is an organization established by CANARIE Inc.
(www.canarie.ca), who provide the national 
research and educational network in Canada; the C3.ca Association
(www.c3.ca), which is an organization that represents
the user community of High Performance Computing (HPC) facilities
in Canada; and the National Research Council of Canada (NRC)
(www.nrc.ca).

The goal of Grid Canada (GC)
is to foster collaboration on Grid activities in Canada.
GC is responsible for the Canadian Certificate Authority
and has helped coordinate a number of demonstration activities such
as the 1 TB high-speed file transfer project between the TRIUMF 
Laboratory in Vancouver and CERN \cite{igrid}
and a grid test using a number of HPC (shared memory machines) 
computers across Canada.
In addition, GC has helped to establish the Linux Testbed 
that will be described in this paper.

The GC Linux Testbed involves researchers
at the University of Victoria who are members of the BaBar
and ATLAS Collaborations.   In addition, personnel at the
NRC Institute for Information Technology, CANARIE Inc. and
the British Columbia Ministry for Management Services  have also
made significant contributions to the Testbed.

The Testbed currently includes 12 sites in Canada ranging from a
number of sites in British Columbia to those 3000 km away in Ontario.
Typically each site has between 1 and 8 computers.  
All sites are required to run Globus 2.0 or 2.2 \cite{globus}
on their computers except for the Victoria site
which is described in the next paragraph.
In addition, sites must also run an OpenAFS client \cite{openafs}.
Latencies between the two most-distant sites vary between
50 and 100 milliseconds.
The remote sites access the application-specific software 
and, for some tests, read the input and write the output data via AFS.

The Victoria Grid testbed is shown in fig~\ref{uvictb}.  
It consists of two servers for AFS and  Objectivity \cite{objy}.
A head node (grid.phys.uvic.ca)
runs the Globus software with a series of processing
nodes (grid1, ..., grid6) hidden behind a firewall.  
Jobs are distributed to the processing nodes using Condor \cite{condor}
although PBS \cite{pbs} has been tested.
The processing nodes have the ability to access the AFS Server
through the firewall.

Access to the software via AFS by many remote machines can 
cause significant degradation in the CPU utilization to the Server
and subsequently reduce the efficiency of the remote processing machines.
AFS has a faster, more efficient method of reading clones or
read-only volumes.
For example, if a volume is read-only, then the client only has to contact
the AFS server once to cache the volume.
We tested AFS performance using two clones and two replicated
volumes  on two separate machines.
The AFS Server machine (basilisk.phys.uvic.ca) held the master and 
a cloned volume, with the Objectivity Server machine (objy.phys.uvic.ca)
holding a replicated and a cloned volume.
This configuration appeared to be sufficient for the size of the
Testbed so that no bottlenecks occurred as a result of the 
AFS Servers.

Initial tests of the ATLAS software using a local machine
and a machine 1200 km distant showed that using a cloned 
volume increased the access to the software at the remote machine.
Real execution time at the remote machine dropped by about 40\% 
whereas (as expected) no change was observed at the local machine.
Running simultaneous multiple jobs over the Grid did not degrade the AFS
performance.   
We concluded that having cloned volumes significantly increased
the AFS performance.    
We plan to continue to monitor the AFS performance as we continue
to scale up the size of the GC Testbed.

Jobs were submitted from Victoria using simple scripts.
No resource broker or portal was used however, we are examining
the available tools (such as GridPort \cite{gridport})
and plan to incorporate them into the Testbed.
Monitoring of the GC Testbed was done using the Cacti package 
\cite{cacti} providing network traffic and CPU utilization.

Detailed descriptions of the Victoria Testbed can be found at 
http://grid.phys.uvic.ca.

% .... Description of Tests

\section{Description of Tests}

HEP simulation applications begin by the generation of events 
which are collections of 4-vectors representing the 
trajectories of particles from an initial collision of two
incident particles.
The 4-vectors are then passed through a program that simulates
the response of the experimental detector.  
The output is in a format that is usually identical to the format
used to store the real collision data, but includes additional
simulation information.
After the detector simulation, the data is passed through an 
application that reconstructs the data into physical quantities
such as tracks in the tracking chambers and energy in the 
calorimeters.
It is now common practice to inject background events (either real
or simulated) on top of the simulated physics events.

We chose to examine the performance of the GC Testbed using
the simulations of the BaBar and ATLAS experiments.
The main difference between the ATLAS and BaBar simulation 
applications is the manner in which they store their data.
ATLAS uses conventional files in a Zebra format developed at CERN
while BaBar uses the commercial Objectivity database.  
In the following subsections we describe the 
results of running the BaBar and ATLAS simulations
on the GC Testbed.
It is worth pointing out that
neither the BaBar or the ATLAS simulation 
are optimized for operation over a wide-area Grid.

\subsection{Tests using the BaBar Monte Carlo Simulation}

The BaBar Collaboration 
studies electron-positron collisions at the SLAC facility. 
In addition to collecting data from the experiment, the collaboration
generates a significant amount of Monte Carlo simulated data in 
order to model the response of the detector and to help assess efficiencies
and systematic errors in the physics studies.

The simulation application is run in 3 phases for event generation,
detector simulation and reconstruction.
Data is written to Objectivity by the event generation application.
The detector simulation reads the output of the event 
generation and writes output that is then used by the reconstruction.
The reconstruction application writes out data that can be used
for physics studies.

The BaBar application accesses the database multiple times per event.
The handshaking required between the database and application combined
with the large latency resulted in relatively poor utilization of
the remote processors.
For example, we observed that CPU utilization was 
approximately 10\% at sites 3000 km away from the Objectivity Server.

\begin{figure*}
\includegraphics[width=130mm]{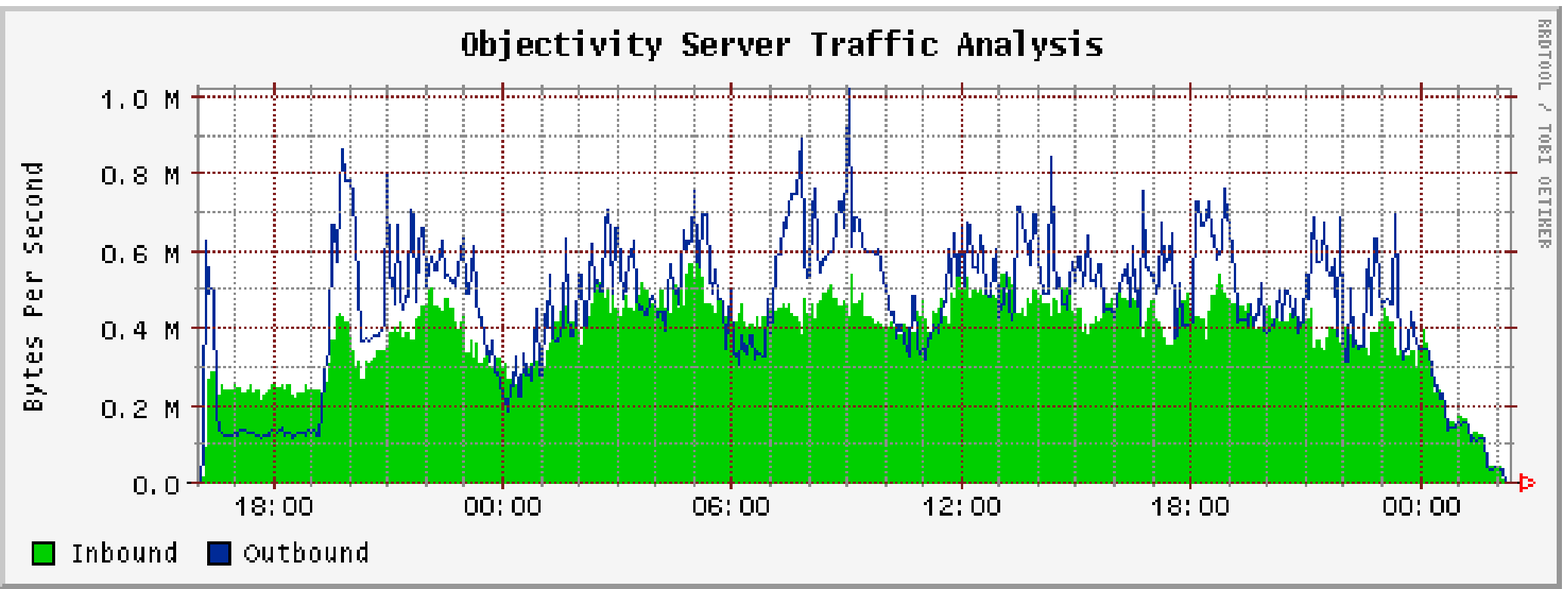}
\includegraphics[width=130mm]{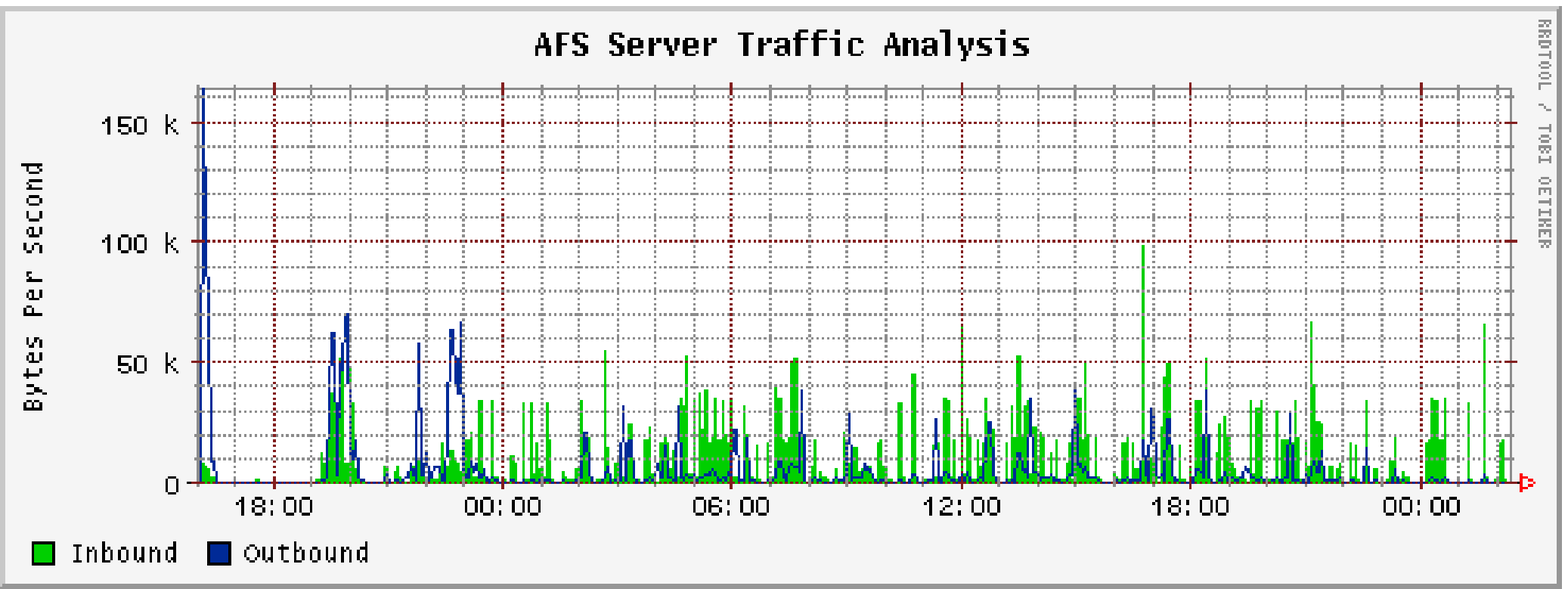}
\caption{Inbound and outbound network traffic to the Objectivity 
(upper plot) and 
AFS Servers (lower plot)
for the BaBar Simulation Application Test.}
\label{f1}
\end{figure*}

A series of 1 day tests were run involving many of the GC Testbed sites.
A total of approximately 50,000 events were generated per test.
The BaBar software was accessed from the AFS Server.
The network traffic to the Objectivity and AFS Servers is shown
in figure~\ref{f1}.
Approximately 0.5 MB/s of data was constantly read and written
to the database.
AFS traffic was generally only a few KB/s with a peak during the
start of the application.

Our initial tests showed that one of the areas that had to be
addressed was the issuing and releasing of database locks.
For example, a remote application locks the entire 
federation during a creation of a container.
If a remote machine lost network connectivity 
while a global database lock was present
then this resulted in the entire Testbed being halted
as none of the sites would be able to write to the database.
As a result we developed a series of monitoring and cleanup
utilities that would eliminate any problems associated with the 
database locks.

We consider the test a success in that we could run the BaBar
application on non-HEP resources.    
It is clear that the efficiency at the remote sites was low
and we believe a number of modifications to the BaBar application
could improve the throughput.
However, given the plan by BaBar and the LHC experiments to move
away from the Objectivity database, it was decided not to pursue
this application further.

\subsection{Tests using the ATLAS Data Challenge software}

The ATLAS Collaboration will study
high energy proton-proton collisions at the Large Hadron Collider 
at the CERN Laboratory.
ATLAS is expected to produce large volumes of data starting in 2007.
The ATLAS Data Challenge (DC) is a project to test and develop
the ability to exploit remote computing resources for Monte
Carlo Simulation.
For our tests, we used the application that generates background 
(called pile-up) events.
Input and output data are standard files in a Zebra format.
Future ATLAS DC will use the Root data format\cite{root}; 
such a change will make no difference to the way in which the current 
test was performed.

We tested three modes of operation on the testbed: 
\begin{enumerate}
\item
Both input data and output data transferred using AFS
\item 
Input data transferred using GridFTP and output data transferred using AFS
\item
Both input and output data transferred using GridFTP
\end{enumerate}
All software and log files use AFS.

It has been recognized that AFS was not an ideal protocol for
transfering large data files, however, we felt it was important
to quantify the differences.
We found AFS to extremely slow for reading input data and resulted
in CPU utilizations being less than 5\%.
AFS caches its output; as a result we tested a mode where the input 
was sent via GridFTP to the remote site and let AFS control the output.
The results of this mode of operation were more encouraging, 
however the performance of AFS is comparable to single stream FTP
and using GridFTP is a much more effective solution.

The mode of operation we chose was one where the input files were
tarred and compressed, and sent via GridFTP to the remote site.  
At the remote site, the files would be untarred and uncompressed,
and the job executed.   
At the end of the job, the output files would be tarred and compressed,
and then transferred via GridFTP back to the Victoria disk server.
This was found to be a successful mode of operation.
Access to software via AFS proved to be reliable and efficient.
We chose an application that would run extremely quickly on
the processing nodes (30 minutes on a 2 GHz CPU).
Compression, transfer and uncompression was approximately 20 minutes
for a 2 GB file.
Although the overall utilization of a remote CPU could be considered
to be low if one includes the compression and transfer times,
a more realistic application, such as simulation or reconstruction,
would take hours rather than minutes 
to process data files of the same size.
In this situation, we believe the CPU utilization would
significantly improve.

% ..........................summary

\section{Summary}

We have demonstrated that HEP applications can be run on generic
computing resources with relatively high efficiency.
With (modest) design changes in the applications we expect that
further gains in efficiency would be possible.
Further tests are planned using a sites with significantly larger
resources with the aim of establishing a production level Grid 
resource in Canada.

% ...........................acknowledgements

\begin{acknowledgments}
We would like to thank those people who made available 
resources at their sites.

The support of the Natural Sciences and Engineering Research
Council of Canada, the National Research Council of Canada,
the C3.ca Association and CANARIE Inc. is aknowledged.

\end{acknowledgments}

% ...........................references

\end{document}